\documentclass{raa}
\usepackage{graphicx}
\usepackage{times}
\usepackage{txfonts}
\usepackage{natbib}
\bibpunct{(}{)}{;}{a}{}{,}

\def \ls           {\hbox{L$_{\odot}$}}
\def \ms           {\hbox{M$_{\odot}$}}           
\def \kms          {\hbox{km$\,$s$^{-1}$}}
\def\um{{$\mu$m~}}

\begin{document}

\title{Properties of the UCHII region G25.4NW and its associated molecular cloud}

\volnopage{Vol.0 (200x) No.0, 000--000}      
\setcounter{page}{1}

\author{M. Ai\inst{1}
              \and M. Zhu\inst{1}
              \and L. Xiao\inst{1}
}

\offprints{L. Xiao}

\institute{National Astronomical Observatories, Chinese Academy of
  Sciences, 20A, Datun Road, Chaoyang District, Beijing 100012,
  China\\ 
  \email{aimei,mz,xl@nao.cas.cn}
}

\date{Received~~2012 month day; accepted~~2013~~month day}

\abstract
{UCHII G25.4NW is a bright IR source in the inner Galaxy region.
New HI images from the VLA Galactic Plane Survey (VGPS) show clear absorption features
associated with the UCHII region up to 95~km~s$^{-1}$, and there
is not any other absorptions up to the tangential velocity.
It reveals that G25.4NW is at a near-side distance of 5.7~kpc, and it is 
located in the inner Galactic molecular ring region.
Using the new distance, the bolometric luminosity of G25.4NW is estimated 
as $10^{5.6}~L_{\odot}$, which corresponds to an O6 star.
It contains 460~$M_{\odot}$ of ionized gas.
High-resolution $^{13}$CO image from the Galactic Ring Survey (GRS)
reveals that G25.4NW is part of a more extended star-formation complex with
about $10^{4}~M_{\odot}$ molecular gas. 
}
$\cdots\cdots$

  
\authorrunning{M. Ai, M. Zhu, L. Xiao \& W. W. Tian }            
\titlerunning{Properties of the UCHII region G25.4NW and its associated
 molecular cloud }

\maketitle

\section{Introduction}
\label{sect:intro}
Massive star formation play an important role in galaxy evolution.
Ultra compact HII (UCHII) regions are indicators of high mass star
formation sites. In these regions, newly born stars are still
embedded in their natal molecular clouds. The energy from the central
massive stars are absorbed by the ambient dust and reradiated in
the infrared (IR), making UCHII regions among the brightest IR sources
in the Galaxy.

G25.4NW is an UCHII region first identified by~\citet{ldw85},
which is found to have the highest FIR luminosity in the inner spiral 
arm region, hosted in a giant molecular cloud (GMC) G25.4$-$0.14. 
It is also on the top two list of IR-luminosity among the 18 radio-selected UCHII
regions studied by~\citet{grb07}. Recently, 
a massive protostar (6-12 $M_\odot$) is found in a massive SCUBA core 
JCMT18354-0649S~\citep{zdw11}, which is about $1'$ south of G25.4NW.
It is believed to lie in the same molecular cloud as G25.4NW based on the smooth change of
$^{13}$CO and $^{18}$CO spectral profiles.
Spectral line studies of JCMT18354-0649S show that both 
infall and outflow motions are seen in the envelope gas that surrounds
this source~\citep{wzw05,lwz11,ckr09}, suggesting that this source is actively accumulating
material and thus can potentially grow to a even more massive object 
as it evolves. It's a middle mass object, and might be in a early state as a
precursor of the UCHII region. 
Hence the G25.4NW and JCMT18354-0649S region is one of
the most active star forming sites in the inner spiral arm region.

However, there was considerable uncertainty about the distance to
the UCHII region G25.4NW. 
~\citet{dwb80} and ~\citet{t79} assigned a far
side distance to G25.4NW based on absorption
features in the H$_{2}$CO~ and OH~ lines at ~96 km/s, which was thought to
be close to the tangent point velocity. ~\citet{srb87} suggested a
far-side distance of 11.5~kpc, based on the scale height of molecular
cloud.  The papers of~\citet{ldw85,cwc90} used the far-side distance of 9.8 kpc for the
UCHII region G25.4NW and they estimated a far-IR luminosity of
$L_{\rm FIR}= 10^{6.238} $~\ls, indicating the presence of an O4 or O5
star. However,~\citet{grb07} found that the mid-IR spectral
type of G25.4NW was consistent with a B1 or an O9-O7 star (based on
the mid-IR dust temperature, with a $L_{\rm MIR} = 10^{4.3} -10^{5.2}
$). This suggests that the stars are not as hot as indicated by the
far-IR data. 

In the case of JCMT18354-0649S,~\citet{zdw11} also found that if
the far-side distance of 9.6~kpc was used, the mass of the central
protostar derived from SED modeling would be as high as 10-20~\ms, and the
temperature would be 4000-10,000~K, equivalent to a B2 star or
earlier. Such a star should have detectable radio continuum and
strong 24~\um MIR emission, which is in contradiction with
observational evidences.
On the other hand, the near-side distance of 5.7~kpc results in more reasonable physical
parameters for JCMT18354-0649S.

Early exploration have mixed the emission and absorption spectra of G25.4NW
with the nearby giant HII region W42~\citep{bcd00} towards this direction due to low resolution. 
And the methods of using the OH absorption feature and scale height method carry a large uncertainty.  
~\citet{ab09} has used the HI Emission/Absorption method with high resolution VGPS and SGPS
HI survey data to resolve the kinematic distance ambiguity for a inner Galaxy HII regions
sample. In the catalogue, they quoted a near-side distance of 5.9~kpc for G25.4NW. 
In this paper, we present the HI absorption feature associated with the UCHII region
G25.4NW to confirm the near-side distance result, and give new estimates to the
related physical parameters, such as the FIR luminosity and spectral type of the massive
stars in G25.4NW. Using GRS $^{13}$CO (1--0) data, we also discuss the molecular cloud environment
of the UCHII region.

\section{Data }
\label{sect:Data}
The radio continuum at 1420~MHz and HI emission data sets were obtained from
the Very Large Array (VLA) Galactic Plane Survey (VGPS), described in
details in~\citet{std06}.  The continuum images of G25.4NW shown in
this paper have a spatial resolution of 1$\arcmin$ at 1420~MHz and an
rms sensitivity of 0.3~K ($T_{b}/S_{\nu}=168$~K Jy$^{-1}$).  The
synthesized beam for the HI line images is also 1$\arcmin$, the radial
velocity resolution is 1.56~km~s$^{-1}$. The short-spacing information
for the HI spectral line images comes from additional observations
performed with the 100~m Green Bank Telescope of the NRAO.

\begin{figure*}[!hbt]
\begin{center}
\includegraphics[angle=0,width=0.48\textwidth]{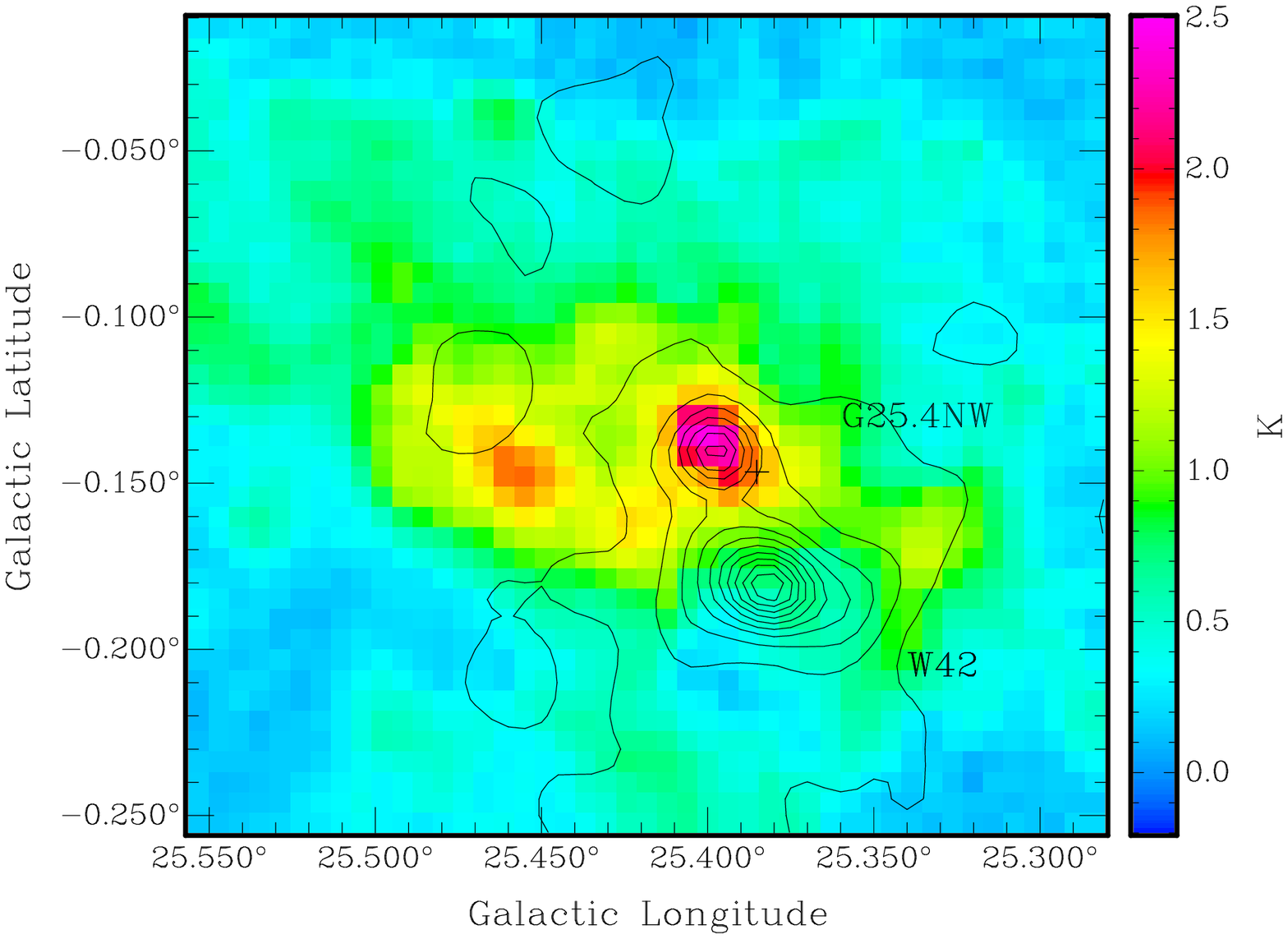}
\includegraphics[angle=0,width=0.48\textwidth]{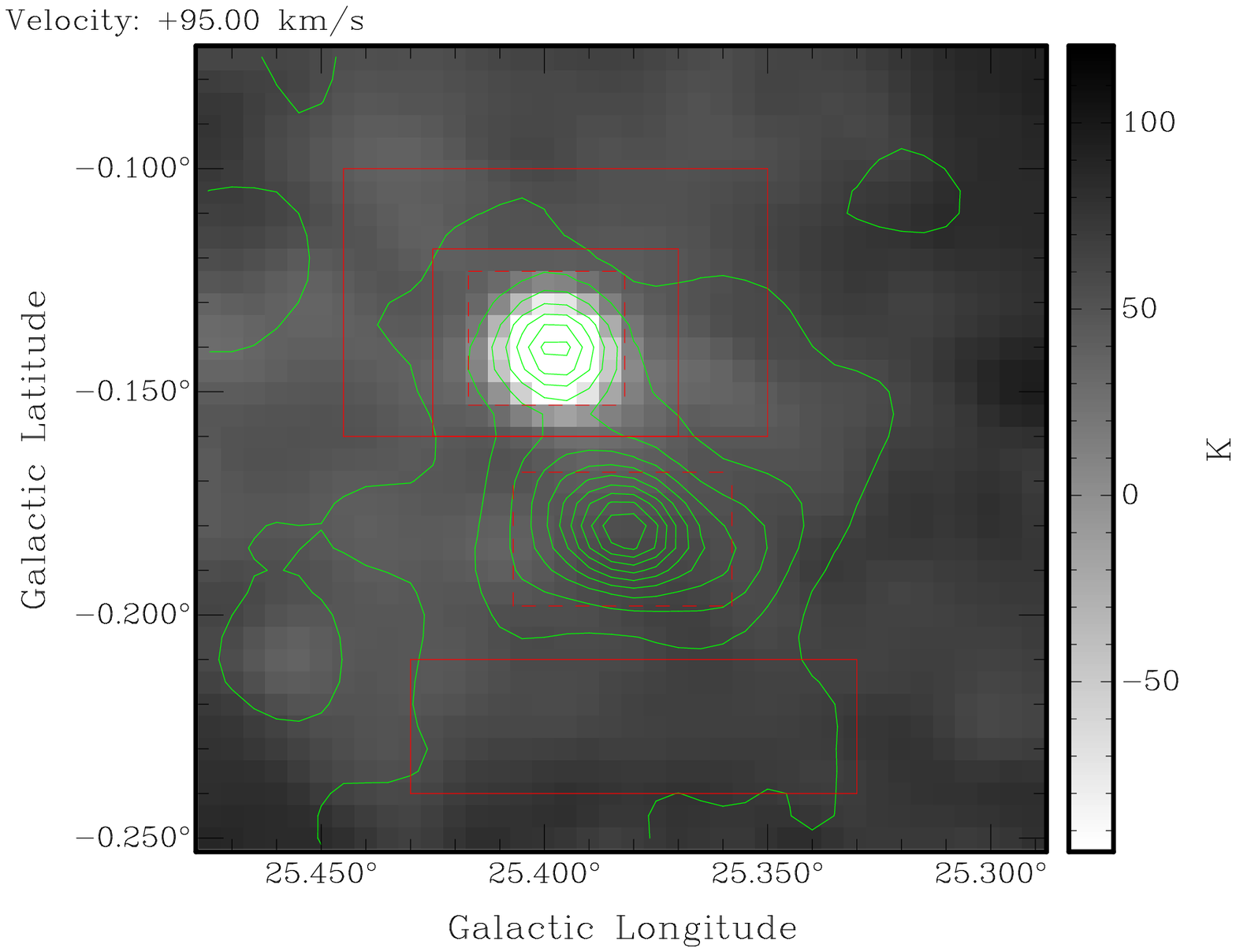}
\caption{The VGPS 1420~MHz continuum contours of G25.4NW overlaid on
(left) the $^{13}$CO integrated intensity image (over the range 90-101~\kms)
and HI single-channel map (right) at 95~km s$^{-1}$ for the GMC G25.4-0.14 region. 
Contours of the 1420~MHz continuum emission of the 3$\times \sigma$ 
($\sigma=1$~K) level above the background emission of 29~K,
and then starting from 67 K in steps of 57 K are overlaid on it. 
The plus sign marks the position of the massive protostar JCMT18354-0649IRS1a
found by~\citet{zdw11}.  }
\label{cont}
\end{center}
\end{figure*} 

The $^{13}$CO-line(J=1$-$0) data sets are from the Galactic ring
survey~\citep{jrs06} carried out with the Five college Radio Astronomy
Observatory 14~m telescope.  The CO-line data used in this paper have a
velocity coverage of $-$5 to 135~km~s$^{-1}$, an angular resolution of
45$\arcsec$ with 22$\arcsec$ sampling, a radial velocity resolution of
0.21~km~s$^{-1}$, and an rms noise of $\sim 0.13$~K.

The Spitzer 8.0$\mu$m images are taken from the Galactic Legacy Infrared
Mid-Plane Survey Extraordinaire (GLIMPSE;~\citet{bcb03}) using the Infrared Array
Camera (IRAC)~\citep{fha04}. The IRAC data were processed by the GLIMPSE team;
image mosaics were created using the GLIMPSE pipeline after artifacts such as
cosmic ray, stray light, column pull-down, and banding had been removed.

\section{Results and analysis}
\label{sect:Results}
\subsection{Continuum images}

Fig.~\ref{cont}(left) shows the VGPS 1420~MHz continuum contours overlaid
on the $^{13}$CO integrated intensity image (over the range 90-101~\kms)
for the G25.4-0.14 region.
Two radio sources, separated by about $2'$,  are marked in the map, namely
G25.4NW (l=25.40\degr, b=-0.14\degr)at and  W42 (l=25.38\degr, b=-0.18\degr).
The later is also named as G25.4SE in~\citet{ldw85}.  A massive
protostar JCMT18354-0649IRS1a was recently discovered at about $1'$
south of G25.4NW~\citep{zdw11}, which is marked as a plus sign.
Contours of the 1420~MHz continuum emission starting from 67~K in
steps of 57~K define a roughly circular region with a diameter of
$\sim1.2\arcmin$. As shown in Lester et al. 1985, G25.4NW has a size
of about $25''$ which is not resolved in this map.

\subsection{The HI absorption spectra and kinematic distance}

We have searched the VGPS data with the radial velocity ranging from
$-$113 to 165~km~s$^{-1}$ for features in HI absorption which might be related
to the morphology of G25.4NW. There are HI absorption features towards
the direction of the HII region
at the following velocity: 5, 20, 40, 50, 55, 67, 95~km~s$^{-1}$.
These HI absorption features towards the continuum intensity of
G25.4NW are caused by absorption in HI clouds between the HII region and the earth.
Figure~\ref{cont} (right) shows the map of HI absorption at the channel of
the highest absorption velocity of 95~km~s$^{-1}$. The map has superimposed contours
of 1420~MHz continuum emission which mark the HII regions. 

We construct the HI emission and absorption spectra of G25.4NW in the same way as
developed by~\citet{tl08,ab09}. To minimize the potential difference in the HI
distribution along the two lines of sight (on-source and background), 
a background region is chosen near the continuum peak, as shown in Fig. 1 (right).  
Assuming that the spin temperature and optical depth of the background region are
similar to that of the on-source region,
the resulting difference in brightness temperature at velocity $v$ is given by:
$\Delta T(v)=T_{\rm off}(v)-T_{\rm on}(v)=T_{\rm c}(1-e^{-\tau(v)})$. Here $T_{\rm on}(v)$ and $T_{\rm off}(v)$ are the average HI brightness
temperatures of the HII region and the adjacent background region from a
selected area.
$T_{\rm c}$ is the average continuum brightness temperatures of the HII region. 
$\tau(v)$ is the HI optical depth from the continuum source to the observer.

\begin{figure*}[!hbt]
\begin{center}
\includegraphics[angle=-90,width=0.48\textwidth]{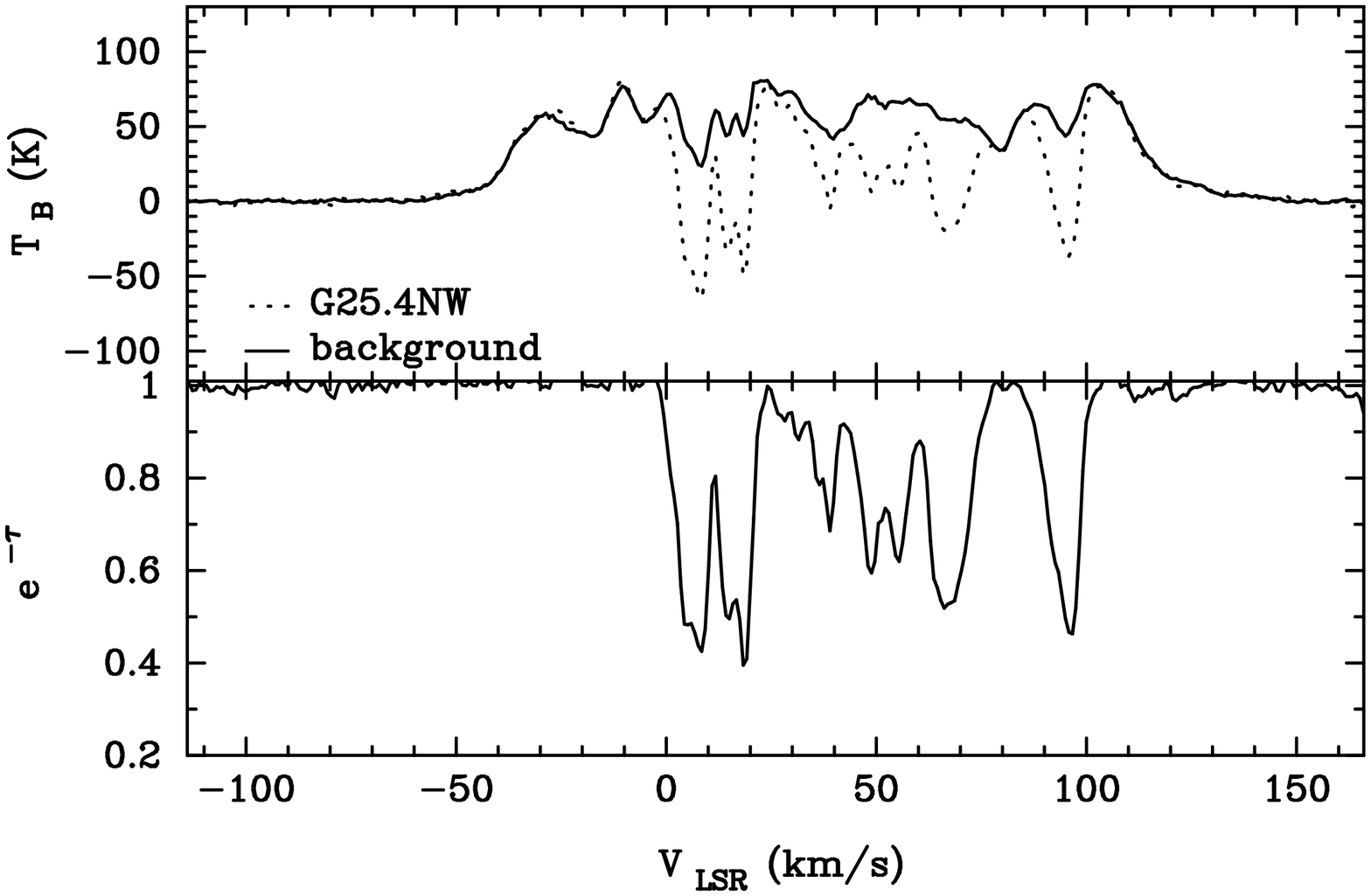}
\includegraphics[angle=-90,width=0.48\textwidth]{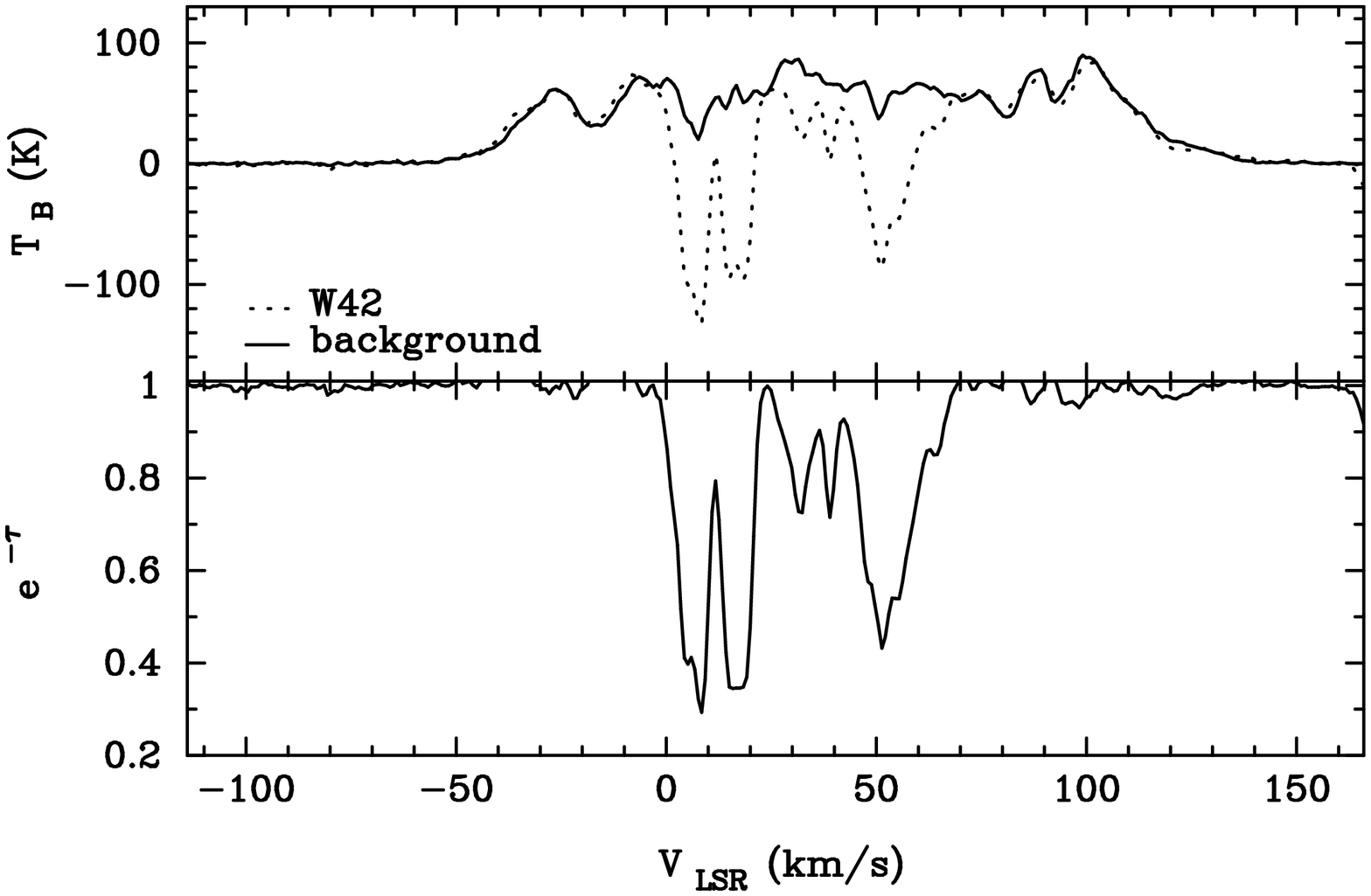}
\includegraphics[angle=-90,width=0.46\textwidth]{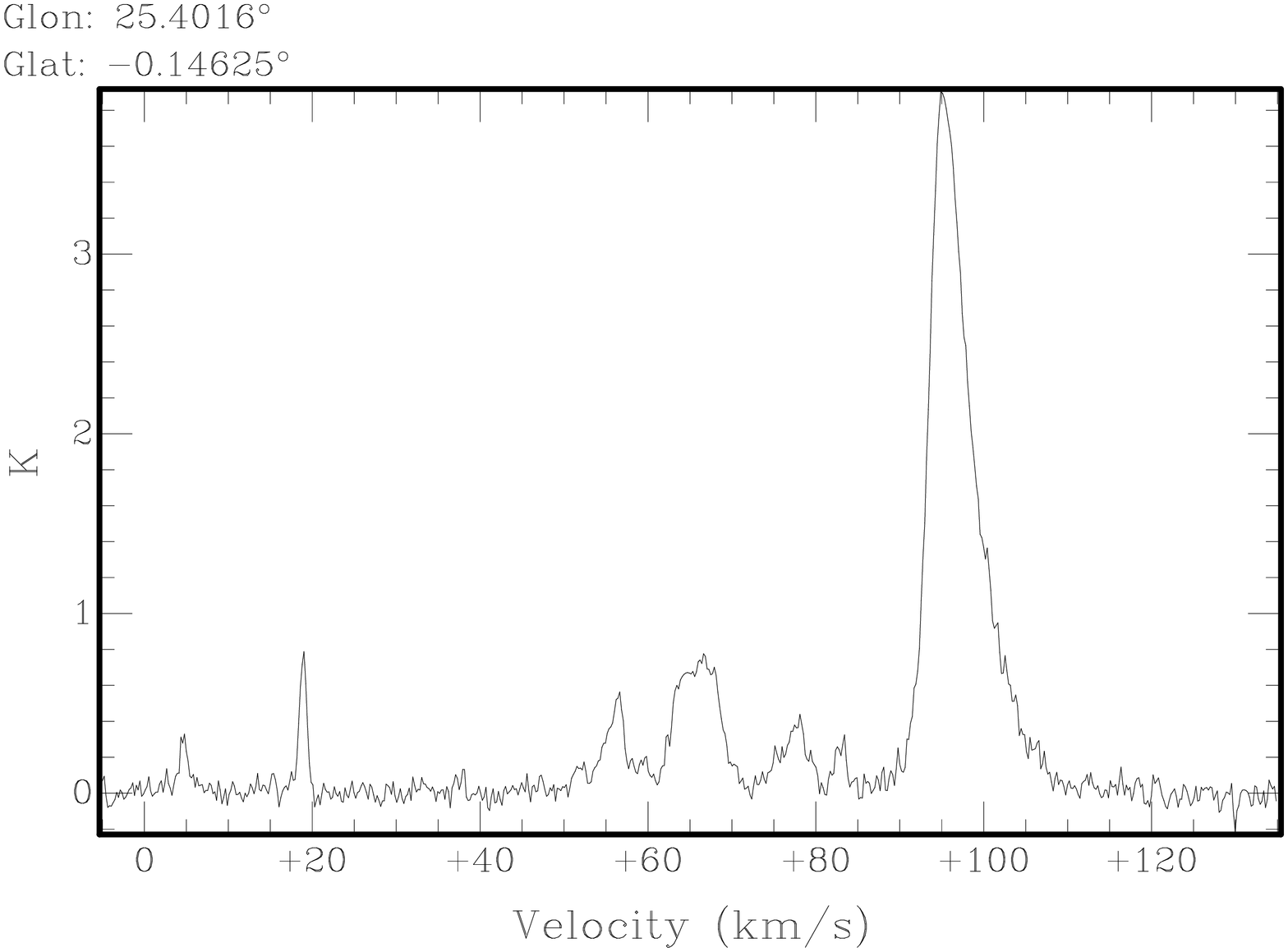}
\includegraphics[angle=-90,width=0.46\textwidth]{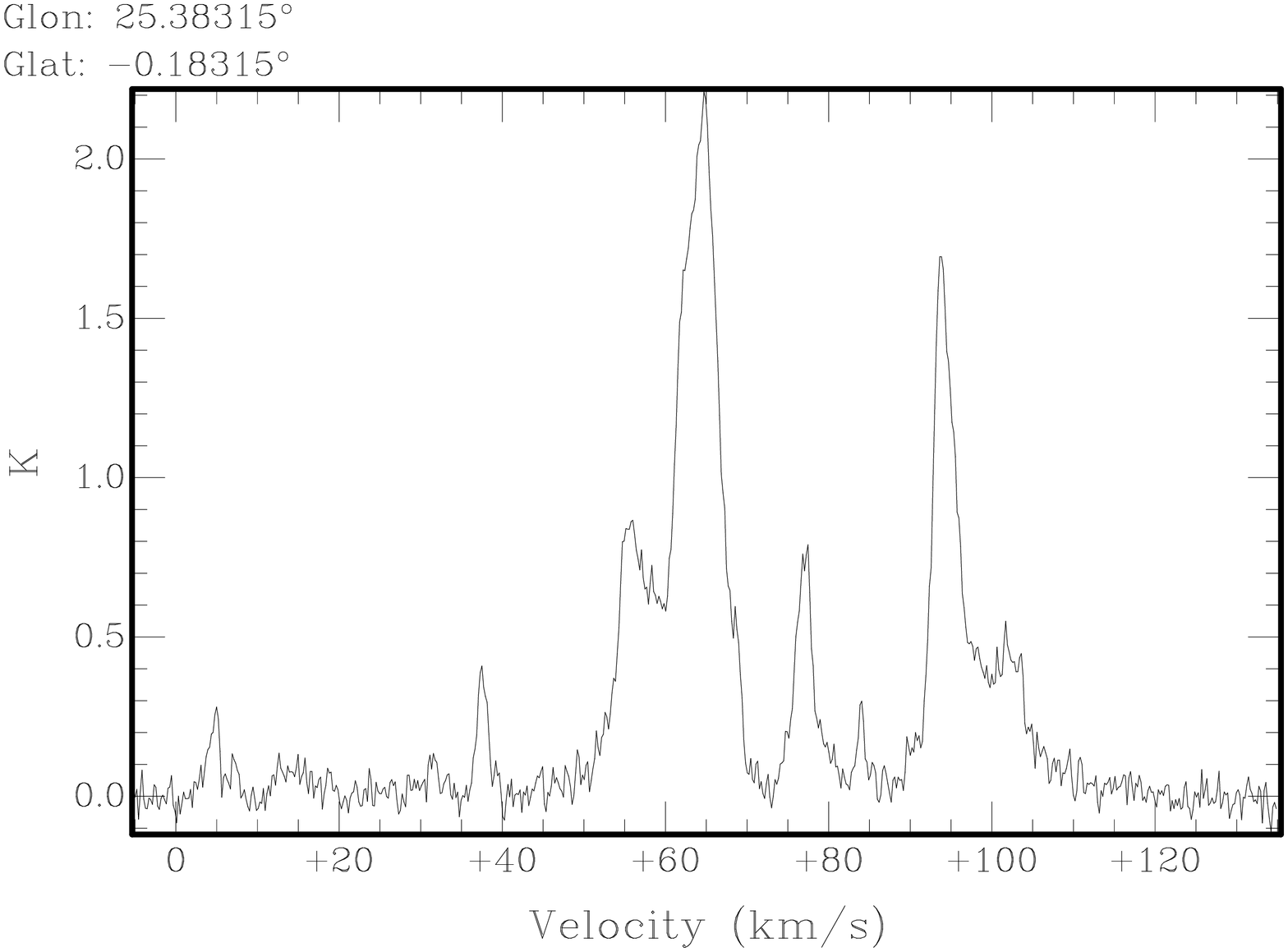}
\caption{Upper panels: HI emission and absorption spectra extracted from the boxes
    around G25.4NW (left) and W42 (right) indicated in Fig.~\ref{cont}.
    Lower panels: Averaged $^{13}$CO emission spectra of the G25.4NW (left) and
    W42 (right) region. The 90-101~\kms emission peak in W42 $^{13}$CO spectra
    comes from the G25.4NW molecular cloud due to the box selection.
}
\label{spec}
\end{center}
\end{figure*}

Figure~\ref{spec} (upper left panel) shows the source and background
HI emission and the absorption spectra for the selected
region on top of G25.4NW. The $T_{\rm on}$ is obtained by averaging over
the region outlined by a dash-line box (Fig 1 right).  
The respective off-source region is defined by the region between two
solid-line boxes, within which the $T_{\rm off}$ is
obtained. 
The averaged continuum temperature of G25.4NW is 154~K.
Figure~\ref{spec} clearly reveals strong absorption
features at 5, 20, 40, 50, 55, 67, 95~km~s$^{-1}$.
The highest absorption velocity detected is close to the tangential velocity
of 120~km~s$^{-1}$, but there is not any absorption feature between them,
which indicates a near-side distance.
Taking a circular Galactic rotation curve model, and using the latest
estimates of the model parameters, i.e. $V_{0}$=220~km~s$^{-1}$and
$R_{0}$=8.5~kpc~\citep{fbs89}, we can determine the kinemtic distance
using the distance-velocity curve as illustrated in Figure 3.  The
highest absorption velocity feature at 95 ~km~s$^{-1}$ corresponds to
a near-side distance of 5.7~kpc.

For comparison, we also produce the HI absorption spectrum for the
selected region of W42, as shown in the upper right panel of
Figure~\ref{spec}. 
The averaged continuum temperature of W42 is 225~K.
The highest HI absorption feature
lies at 50$-$65~km~s$^{-1}$. It is far away from the tangential
velocity of 120~km~s$^{-1}$, indicates also a near-side distance of
4.6~kpc. The 90-101~\kms emission peak in W42 $^{13}$CO spectra
comes from the G25.4NW molecular cloud due to the box selection,
and high resolution observation will solve this problem.

\begin{figure}[!hbt]
\begin{center}
\includegraphics[angle=-90,width=0.42\textwidth]{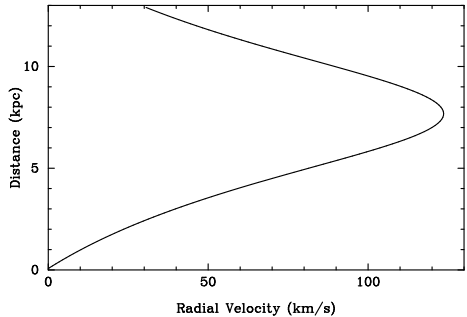}
\caption{The relation between the distance and radial velocity for the
    assumed rotation curve.
}
\label{dist}
\end{center}
\end{figure}

\begin{table*}[!htbp]
\caption{Summary of HI and CO velocity features}
\label{arm}
\begin{center}
\begin{tabular}{cccccccc}
\hline\hline
 Crossing number       &local & 1     & 2   & 3   & 4   & 5   & 6   \\
\hline
Central velocities of the model spiral arms (km~s$^{-1}$)      &  0   & 26    &71   &93    &52   &21   & 0  \\
G25.4NW HI absorption feature&0-10  &10-20  &60-80&90-100&  &   &    \\
G25.4NW CO emission feature  &5     &15-20  &75-85&90-100&50-60&15-20&5  \\
\hline
W42 HI absorption feature&0-10  &15-20  &50-65&      &     &     &    \\
W42 CO emission feature  &5     &32     &60-70&90-100&50-60&15-20&5  \\
\hline
\end{tabular}
\end{center}
\end{table*}

The optically thin $^{13}$CO emission could clearly reveal the distribution of
molecular clouds along the direction of the HII region, and will indicate possible origin
of these HI absorption features.
Figure~\ref{spec} (lower left panel) shows an average
$^{13}$CO spectrum for the full velocity range in the direction of
G25.4NW.  It shows 7 molecular components with high brightness
temperature, at radial velocities of 5, 18, 55, 65, 78, 83,
95~km~s$^{-1}$ respectively. Most of the $^{13}$CO emission peaks appear to
be associated with an HI absorption feature, and the $^{13}$CO emission
velocities are consistent with the spiral arm radial velocities of the 
Milky Way based on the Galactic arms scheme~\citep{cl02}.  In
Table~\ref{arm} we list the respective spiral arm radial velocities
and possible associations with $^{13}$CO emission or HI absorption features.
Cases are included even when the velocities are up to
$\sim$10~km~s$^{-1}$ different, since the calculated arm crossing
velocities are based on a circular rotation model and can be off by
$\sim$10~km~s$^{-1}$ due to non-circular motions from spiral arm
shocks and velocity dispersion.

In summary, both W42 and G25.40-0.14 are at the near side distance,  and they
appear to be located in the inner galactic CO ring at 4-6 kpc.

\section{Discussion}
\label{sect:discussion}
\subsection{Infrared emission}
Dust associated with HII regions can absorb radiation from the
exciting stars of the nebulae and re-emit in the infrared wavelengths. 
If there is
sufficient spatial coverage and optical depth of dust surrounding an
HII region, measurement of the infrared flux from the HII region can
provide information on the bolometric luminosity ($L_{bol}$) of the
exciting stars.

\begin{figure}[!hbt]
\begin{center}
\includegraphics[angle=-90,width=0.45\textwidth]{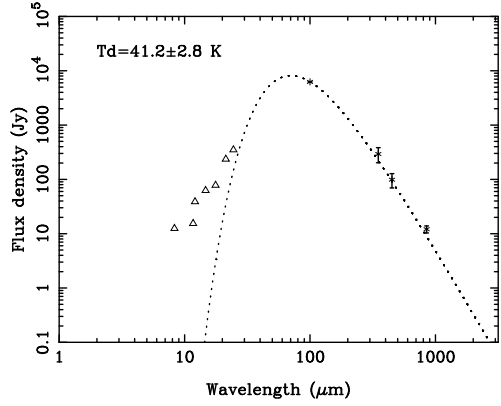}
\caption{Flux density distribution of G25.4NW fitted with a one-component modified blackbody
model. The line marks the best fit to the 100, 350, 450, 850$\mu$m flux densities.
}
\label{Lbol}
\end{center}
\end{figure}
 
We collect the total MIR flux densities of the IR counterparts
for G24.5NW in the literature, 
including data from the \textsl{MSX} point-source catalog (8.3, 12.1, 14.7,
and 21.3~$\mu$m), Keck I 10~m telescope (11.7, 17.65,
24.5~$\mu$m)~\citep{grb07}, IRAS 100~$\mu$m~\citep{ldw85}, CSO (350~$\mu$m)
and SCUBA (450~$\mu$m, 850~$\mu$m)~\citep{wzw05}.  \textsl{MSX} had a
beam of $18.3\arcsec$, and the KecK's data had an aperture of $10.5\arcsec$.
The  100~$\mu$m's beam is $50\arcsec$, which is much larger than that
of the CSO and SCUBA data.
We used an aperture of $50\arcsec$ to derive the sub-millimeter fluxes in order
to cover the whole HII region.  The dust emissivity index ($\beta$)
has been estimated to be a typical value of 2.0~\citep{zdw11}. Thus
we used a modified single temperature blackbody curve,
$\nu^{2}B_{\nu}(T_{d})$ to fit the data and derive the spectral engergy distribution (SED). 
In lieu of a radiative transfer model, we have
chosen to fit the cold component of the emission (100$-$850~$\mu$m)
because it contains the bulk of the source luminosity. The SED fitting
is plotted in Fig.~\ref{Lbol}. The best fit parameter is
$T_{d}$=42.5$\pm$3~K.  It is larger than that estimated in
~\citet{zdw11}, considering the contribution of MIR data in
the SED fitting, which trace the central bright temperature component.

\begin{table}[!htbp]
\caption{Summary of observed flux densities of G25.4NW }
\label{sed}
\begin{center}
\begin{tabular}{c r@{.}l r c}
\hline\hline
 Wavelength &\multicolumn{2}{c}{Flux} & error & aperture   \\
($\mu$m)    &\multicolumn{2}{c}{(Jy)} & (Jy)  & ($\arcsec$)  \\\hline
8.3         & 12&51  & --      &18.3  \\
12.1        & 38&86  & --      &18.3   \\
14.7        & 62&48  & --      &18.3    \\
21.3        &235&88  & --      &18.3    \\
11.7        & 15&4   &  0.5     &10.8      \\
17.7        &  78&0  & 3        &10.8   \\
24.5        & 350&0  &80        &10.8    \\
100         &6240&0  & --      &50     \\
350         &293&0   &91       &50    \\
450         & 98&7   &29       &50    \\
850         & 12&1   &1.8       &50   \\  
\hline
\end{tabular}
\end{center}
\end{table}

For a distance of 5.7~kpc we obtain log($L_{bol}/L_{\odot}$)=5.6.
This value is lower than the result of \citet{cwc90} which was
estimated from the far-IR
data using a far-side distance.
Thus the central stars are not as hot as
estimated before, and an  O6 star is expected for such bolometric luminosity.
~\citep{p73}.  

\subsection{Radio emission}
Radio continuum measurements of HII regions can provide a lower limit
to the total number of ionizing photons ($N_{L}$) and estimates of the
average excitation parameter ($U$). Assuming an optically thin,
spherical, constant-density HII region, the equations used
are as follows~\citep{rbg96},
\begin{equation}\label{NL}
N_{L}=7.5\times10^{43}S_{\nu}d^{2}\nu^{0.1}T_{e}^{-0.45}~\rm  [ s^{-1} ]
\end{equation}
\begin{equation}\label{U}
U=1.33S_{\nu}^{1/3}d^{2/3}\nu^{1/30}T_{e}^{0.116}~\rm  [ pc~cm^{-2} ]
\end{equation}
where $\nu$ is the frequency in GHz, $S_{\nu}$ is the flux density
measured at frequency $\nu$ in mJy, $d$ is the distance to the source
in kpc, and $T_{e}$ is the electron
temperature in units of $10^{4}$~K. Parameter values used are
$\nu$=1.42~GHz, $d=5.7$~kpc, and $T_{e}=10000$~K. A general value of
$10^{4}$ K is used because the results of both $N_L$ and U are not
sensitive to $T_{e}$,

The flux density for G25.4NW obtained from the VGPS continuum image
is $1.35\pm0.2$~Jy, which is in agreement with~\citet{grb07} from an VLA
B configuration 1.4~GHz survey with a resolution of $\sim 5\arcsec$. 
This yields a value of
$N_{L}=(3.4\pm0.2)\times10^{48}$~s$^{-1}$, 
$U=47.5\pm10$~pc~cm$^{-2}$. 
Comparing these values (log$N_{L}$=48.5)
to that of ~\citet{p73}, we estimate an ionizing flux and excitation parameters
corresponding to an O6$-$O7~ class star. From the 3-color HKL$\arcmin$
band image (Fig. 3 in~\citet{zdw11}), it is more likely that the
ionizing flux comes from several stars.  
It possible that G25.4NW is powered by serveral O/B stars, with at least one
earlier than an O7 star.

\subsection{Ionized Gas Mass}
For optically thin emission, the emission measure ($EM$) is related to
the surface brightness ($I_{\nu}$) according to the following 
formula~\citep{o89}:
\begin{equation}\label{Tb}
I_{\nu}=2\nu^{2}kT_{b\nu}/c^{2}=j_{\nu}\int n_{i}n_{e}ds=j_{\nu} EM,
\end{equation}
where $j_{\nu}$ is the free-free emissivity for H$^{+}$
(3.45$\times10^{-40}$~ ergs~cm$^{-3}$s$^{-1}$Hz$^{-1}$sr$^{-1}$ at
1420~MHz and 7500~K; after Lang~1980), $n_{i}$ is the ion density, and
$n_{e}$ is the electron density. For $j_{\nu}$ of a pure hydrogen
nebula one can write
\begin{equation}\label{E}
EM =5.77\times10^{2}T_{b\nu}(\frac{T_{e}}{7500~K})^{0.35}(\frac{\nu}{1.42~GHz})^{2.1}~\rm [ cm^{-6}~pc ]
\end{equation}
\begin{equation}\label{E}
T_{b\nu}= \frac{c^2 S_\nu 10^{-26}}{2 k\nu^2 \Omega} [ K ]
\end{equation}
where $T_{e}$ is the electron temperature in the nebula, $S_\nu$ is
the integratged flux density in Jy and $\Omega$ is the radio source
solid angle in steradians.
The UCHII region was not resolved in the VGPS map (resolution 1
arcmin), but it is resolved in the high resolution map observed with the
VLA B configuration~\citep{gbh05}, from which we estimated that
the souce has a size of $10.1''$, corresponding to a physical size of 
$\sim$0.28~pc. Hence we derive
$EM =5.5\times 10^{6}$~cm$^{-6}$~pc using a typical value $T_{e}=10000$~K.
The electron density in the HII region can be calculated from the 
emission measures using the following formula~\citep{grb07}
\begin{equation}\label{E}
n_e= 990 \sqrt{ (\frac{EM}{10^6 pc~cm^{-6}})(\frac{pc}{D})}  [cm^{-3}]
\end{equation}
where $D$ is the diameter of G25.4NW ($\sim 0.28$~pc). 
We find an average value of $n_e=4.4\times10^3~\rm cm^{-3}$.  Assuming a
spherical volume of ionized gas, we estimate G25.4NW contains
460~$M_{\odot}$ of ionized gas.

Comparing the $EM$ and $n_e$ values with those in~\citet{grb07}
(after correcting the distance from 9.6 kpc to 5.7 kpc), our derived
values are slightly lower than theirs. This is because we used a
larger source size. ~\citet{gbh05} used the 5GHz VLA data to estimate
a source size of 3.7\arcsec, thus they obtained values for a more compact
HII region. As the NIR map of~\citet{zdw11} has shown that there are 
at least 3 stars in the G25.4NW region, the VLA map at 1.4GHZ could not
resolved the individual HII region, thus we could overestimate the source
size. Nevertherless, our estimated values of $EM$ and $n_e$ can be used 
as a lower limit. 

\begin{figure*}[!hbt]
\begin{center}
\includegraphics[angle=0,width=0.80\textwidth]{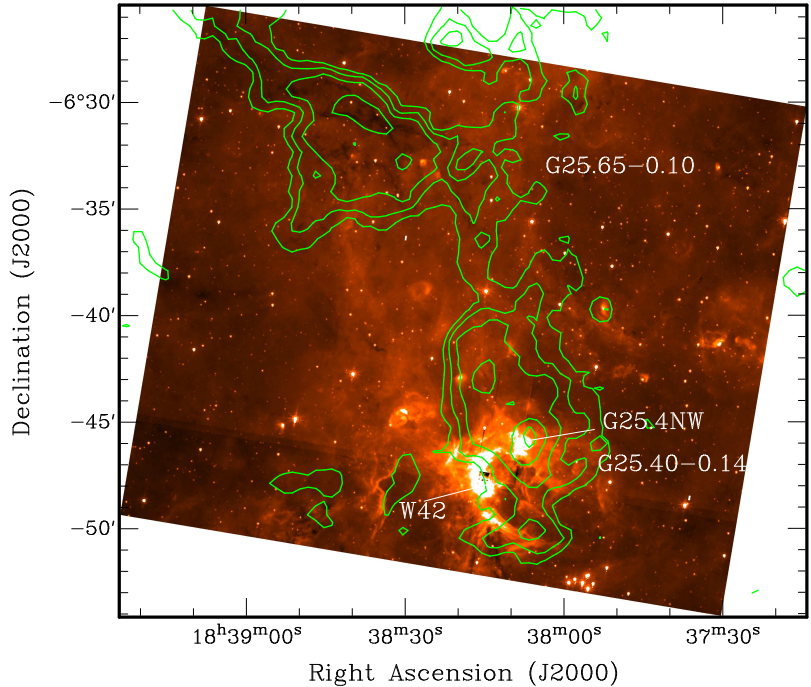}
\caption{The $^{13}$CO integrated intensity contours overlaid on a Spitzer IRAC 8 \um image. The contours are plotted at 0.6 ($3\sigma$), 0.8, 1.1, 1.6, 2.2~K.
}
\label{CO_on_8um}
\end{center}
\end{figure*}

\subsection{The Giant Molecular cloud surrounding G25.4NW}

Figure~\ref{CO_on_8um} shows the $^{13}$CO integrated (over the range
90-101 \kms) intensity contours overlaid on the IRAC 8$\mu$m image. The
MIR 8$\mu$m emission is dominated by the PAH emissions that are heated by
UV photons from young stars. The two bright regions in the map are
the HII regions G25.4NW and W42.  
The W42 region is not physically related with G25.4NW. It is a foreground
source at a different distance of 4.6~kpc (see Sect.3.2).
We can see that
G254.NW is embedded in a giant molecular complex, which appears to be
divided into two major clouds: the fish-shape cloud G25.40-0.14
hosting G25.4NW and the other one G25.65-0.10 that peaks at (l=25.68
\degr, b=-0.13\degr).  The later was named as G25.65-0.10 in the
paper of~\citet{srb87}, which appears to be associated with an IR dark
cloud (IRDC). 
We note that~\citet{srb87} use G25.65-0.10 to refer to
the GMC cloud at 94 km/s, with the size of $13\arcmin\times 8\arcmin$,  
which did not include the G25.4-0.14 cloud. 
However, in our maps these two clouds appear to be associated 
with each other connected with low level contours (as shown in the 3$\sigma$ contour), 
indicating a possible physical connection between G25.4-0.14 and G25.65-0.10  
and they might belong to one GMC at the same distance.
If confirmed, such a GMC would have a maximum physical size of about 36~pc
(22$\arcmin$) at a distance of 5.7 kpc.
The two separated major clouds and the existance of
a number of local peaks in each cloud suggest that 
fragmentation is taking place in this complex, with G25.4NW being the most active
region forming a cluster of massive stars. 
JCMT18354-0649S as another massive core
that hosts a massive protostar is offset from the G25.4NW radio continuum by $1\arcmin$,
indicating that OB stars run away from their natal clouds and the HII region is more
evolved. It appears that the formed massive stars is triggerring new star formation
in JCMT18354-0649S.
We have searched the literature 
and the SIMBA database, and there are no IR or radio sources associated
with G25.65-0.10, suggesting that this cloud is relatively quiet
and no star formation activity is taking place.

The total H$_{2}$ mass of the fish-shape CO cloud G25.4-0.14 (size: 12$\arcmin\times6\arcmin$)
at 95$\pm$10~km~s$^{-1}$ is estimated using $M_{\rm H_{2}}=N_{\rm
H_{2}}\Omega d^{2}(2m_{\rm H}/M_{\odot})$, where $\Omega$ is the solid
angle of the cloud, and $d$ is its distance of 5.7~kpc.
Assuming the CO cloud is under thermodynamic equilibrium (LTE) and the
$^{13}\rm CO$ is optically thin and using the fitted dust temperature of 42.5~K, 
we obtained a $^{13}\rm CO$ column density of the cloud is 
N$_{^{13}\rm CO}=2.91\times10^{17}$~cm$^{-2}$ from the $^{13}$CO integrated map
(over the range 90-101 \kms). 
Then we obtain an average H$_{2}$ column density of
$N_{\rm H_{2}}=4.59\times 10^{5} \rm N_{^{13}\rm CO} \approx1.33\times10^{23}$~cm$^{-2}$, and a molecular cloud mass
of $M_{\rm H_{2}}\approx1.0\times10^{4}~M_{\odot}$. This molecular cloud has a
mean density of $\sim 2.3\times10^{4}$~cm$^{-3}$. 

\section{Summary}
\label{sect:summary}
In Summary, using new VLA continuum and HI-line data for G25.4NW, we
have obtained HI absorption profiles towards the UCHII region.
G25.4NW shows HI absorption features up to 95~km~s$^{-1}$, and there
is not any other absorptions up to the tangential velocity, giving a
near-side distance of 5.7~kpc. The detected HI absorption and CO emission is
coincident with the expected velocity for a galactic molecular ring region
at 4-6~kpc from the Galactic arms model. 
Using the new distance, we obtain a bolometric luminosity
of $10^{5.6}~L_{\odot}$ from collected IR emission from G25.4NW, which correspond to an 
O6 star. We also use the radio
continuum emission data confirm this results, and re-calculate the $EM$ and
electronic density $n_e$.  We estimated that G25.4NW contains
460~$M_{\odot}$ of ionized gas. The $^{13}$CO data reveal that
G25.4NW is part of a more extended star-formation complex with about
$10^{4}~M_{\odot}$ molecular gas.

\begin{acknowledgements}
We acknowledges the supports from NSFC grant 11073028 and also from the 
``Hundred-talent program'' of Chinese Academy of Sciences. We thanks
Prof. Wenwu Tian, Yuefang Wu and Dr. Lei Zhu for reviewing the manuscript
and providing helpful instruction. We thank the anonymous referee for his
instructive and useful comments.

\end{acknowledgements}

\bibliographystyle{raa}

\bibliography{G254}

\end{document}